\documentclass{article}
\usepackage{spconf,amsmath,graphicx,multirow}
\usepackage{mathrsfs}
\usepackage{color}
\usepackage{CJKutf8}
\definecolor{red}{rgb}{1,0,0}
\definecolor{green}{RGB}{20,180,92}
\definecolor{blue}{RGB}{43, 145, 213}
\definecolor{gray}{RGB}{120,115,115}
\definecolor{orange}{RGB}{228,133,9}

\title{SeACo-Paraformer: A Non-Autoregressive ASR System with Flexible and Effective Hotword Customization Ability}
\name{Xian Shi, Yexin Yang, Zerui Li, Yanni Chen, Zhifu Gao, Shiliang Zhang}
\address{Speech Lab of DAMO Academy, Alibaba Group, China}
\begin{document}
\ninept
\maketitle
\begin{abstract}
Hotword customization is one of the concerned issues remained in ASR field - it is of value to enable users of ASR systems to customize names of entities, persons and other phrases to obtain better experience. The past few years have seen effective modeling strategies for ASR contextualization developed, but they still exhibit space for improvement about training stability and the invisible activation process. In this paper we propose Semantic-Augmented Contextual-Paraformer~(SeACo-Paraformer) a novel NAR based ASR system with flexible and effective hotword customization ability. It possesses the advantages of AED-based model's accuracy, NAR model's efficiency, and explicit customization capacity of superior performance. Through extensive experiments with 50,000 hours of industrial big data, our proposed model outperforms strong baselines in customization. Besides, we explore an efficient way to filter large-scale incoming hotwords for further improvement. The industrial models compared, source codes and two hotword test sets are all open source.

\end{abstract}

\section{Introduction}

End-to-end speech recognition has made significant progress in the past decade, with several classic and high-performance ASR backbone models including Transducer~\cite{graves2013speech}, listen-attend-and-spell~(LAS)~\cite{chan2016listen} and Transformer~\cite{vaswani2017attention} receiving considerable attention and spawning numerous variants to address various issues in the field of ASR like streaming ASR~\cite{rao2017exploring,moritz2020streaming,gao2020universal}, multi-lingual ASR~\cite{zhou2018multilingual,karita2019comparative}, and non-autoregressive ASR~\cite{yu2021boundary,gao2022paraformer}, among others. In commercial ASR systems, especially those applied in vertical domains, enabling users to input personal names, place names, named entities and other words as hotwords to obtain personalized recognition results, in short supporting \textit{hotword customization}, is a question of both academic and commercial value.

At the age of conventional ASR systems, as the acoustic model and language model focus on the phonetic and linguistic information separately, the personalized and biased information are introduced by adjusting the weights in weighted finite state transducer~(WFST) decoder~\cite{aleksic2015improved,hall2015composition}. For E2E ASR systems without WFST decoder, several works are explored to enable users to customize their own hotwords~\cite{pundak2018deep,han2021cif,huang2023contextualized}. Contextual listen, attend and spell~(CLAS)~\cite{pundak2018deep} is an E2E contextual ASR modeling method proposed by Golan Pundak et al. They first proposed to introduce a multi-headed attention~(MHA) which is jointly trained with randomly sampled phrases in LAS model. Intuitively, the MHA is trained to capture the relationship between hotword embedding and decoder output in each step. leading the decoder to generate biased probabilities once matched. Such strategy is proved an efficient way for introducing bias information to ASR models without damaging the recognition performance, and then becomes a widely adopted contextual ASR solution.

Building upon the fundamental strategy of CLAS: random hotwords sampling and MHA, several extensions are explored to achieve hotword customization~\cite{han2021cif,huang2023contextualized,hu2019phoneme,jain2020contextual,sathyendra2022contextual,han2022improving}. These methods can be categorized as \textit{explicit} and \textit{implicit} ones. Huang et al. recently proposed contextual phrase prediction network~(CPP Network)~\cite{huang2023contextualized} which is similiar with CLAS (both \textit{implicit}). An extra CTC~\cite{graves2006connectionist} head is introduced for predicting hotwords. By jointly optimizing the CTC module, the encoder output is guided to obtain hotword information thus the ASR results are biased with given hotwords in inference stage.
CPP network can be adopted in CTC, Transducer and Transformer models. Han et al. proposed Continuous integrate-and-fire~(CIF) based collaborative decoding~(ColDec)~\cite{han2021cif}. It is an \textit{explicit} method based on CIF model to achieve hotword customization, predicting an external hotword probabilities which is synchronous with ASR prediction. In the training stage, MHA is conducted between CIF output $\mathbf{E}$ and hotword embeddings. Each step of the MHA output carries hotword attended semantic information $\mathbf{g}$, then a feed-forward layer turns it to probabilities for computing cross-entropy~(CE) loss with the hotword-position-aware target~(e.g. `\textbf{A B C D E F}' for ASR target, an `\textbf{\color{green}{A B}} \textbf{\# \# \# \#}' for bias target when `\textbf{\color{green}{A B}}' is sampled as a hotword and `\textbf{\#}' for no-bias label). The introduced modules actually work as a hotword detector to generate biased probabilities 
 synchronous with ASR predictions. In the inference stage, ASR probabilities and biased probabilities are merged. In~\cite{han2022improving}, the authors propose to filter the incoming hotword list with the attention weight which is proved to be an effective method for further improvement.

However, each of the aforementioned methods has its own respective shortcomings. The vanilla CLAS does not demonstrate consistent effectiveness - the label insertion strategy proposed in \cite{pundak2018deep} is not applicable to all E2E systems. Additionally, implicit modeling make it hard to decouple the general ASR modeling and contextual modeling when the system encounters bad cases. Apparently, comparing to CLAS and its extensions, CIF ColDec has a more intuitive and controllable hotword predicting process and training for the external parameters is decoupled from ASR model. On the other hand, the ASR backbone it employs does not attain the high accuracy as attention-encoder-decoder~(AED) models. Such factors motivate us to propose a novel hotword customization enabled ASR system: \textbf{Se}mantic-\textbf{A}ugmented \textbf{Co}ntextual \textbf{Paraformer}~(SeACo-Paraformer). We adopt Paraformer~\cite{gao2022paraformer} as our backbone model, which serves as a highly efficient NAR model while also maintaining the high accuracy of mainstream AED models. The parallel decoder inside Paraformer allow us to deploy a more complex customization modeling strategy based on CIF ColDec. To address the issue of performance decrease resulting from an expanding hotword list, we use attention score filtering~(ASF) to filter the large scale incoming hotwords. We conduct a series of experiments over large scale industrial data to validate the performace of SeACo-Paraformer and ASF. The paper is structured as follows: Section 2 provides a brief introduction to the adopted CIF and Paraformer. Section 3 presents a detailed introduction to SeACo-Paraformer, including training, inference, and other techniques. Section 4 presents the experimental setup and results, while Section 5 analyzes the performance of the models. Finally, Section 6 concludes the paper.

\section{PRELIMINARIES}

\subsection{Continuous Integrate-and-Fire}

Continuous integrate-and-fire~(CIF) is an alignment mechanism which utilizes the monotonic characteristic of ASR task to predict number of output tokens and obtain acoustic embedding with encoder output. CIF predicts weights for each frame and accumulates them from begin to end. Once the weights added up to 1.0, the encoder output of the former periods are integrated~(weighted sum) to generate a step of acoustic embedding. Considering encoder output $\mathbf{e}_{1: T}$ with predicted weights $\alpha=(0.4,0.8,0.3,0.5,0.2)$,
The integrated acoustic embedding $\mathbf{E}_{1}=0.4\times\mathbf{e}_{1}+0.6\times\mathbf{e}_{2}$, $\mathbf{E}_{2}=0.2\times\mathbf{e}_{2}+0.3\times\mathbf{e}_{3}+0.5\times\mathbf{e}_{4}$. The sum of weights $\alpha$ is $L^{'}$ - the token number of output sequence. The most important features of CIF is that the embedding it generates has the same length as the target sequence, which means it's a entirely acoustic based representation but synchronous with predictions.

\vspace{-3mm}
\subsection{Paraformer}
In this work, we adopt Paraformer~\cite{gao2022paraformer}, a novel NAR ASR model as our backbone. Paraformer stands for \textbf{Para}llel Trans\textbf{former} which contains a decoder receiving acoustic embedding or semantic embedding at the same time in training. Briefly, Paraformer achieves non-autoregressive decoding capacity by utilizing CIF~\cite{dong2020cif} and two-pass training strategy as illustrated in Fig.~\ref{seaco}~(lower dotted box). A CIF predictor is trained to predict the number of tokens and generate acoustic embedding $\mathbf{E_{1:L^{'}}}$ for parallel decoder, which makes up Pass1 in training (w/o gradient). The steps in char embedding $\mathbf{c_{1:L}}$ will be gradually replaced by $\mathbf{E_{1:L^{'}}}$ as accuracy raises in Pass1 and the so called semantic embedding $\mathbf{S_{1:L}}$ is generated according to the correctly recognized positions. Getting rid of the massive computation overhead introduced by auto-regressive decoding and beam-search, Paraformer gains more than 10x speedup with even lower error rate. More implementation details and experiment results can be found in \cite{gao2022paraformer,gao2023funasr}.

%\begin{figure}[htbp]
%    \centering
%    \includegraphics[width=0.4\textwidth]{ts-paraformer.pdf}
%    \caption{Paraformer architecture.}\label{para}
%\end{figure}

\section{SeACo-Paraformer}
\subsection{Architecture and training}

\begin{figure}[]
    \centering
    \includegraphics[width=0.40\textwidth]{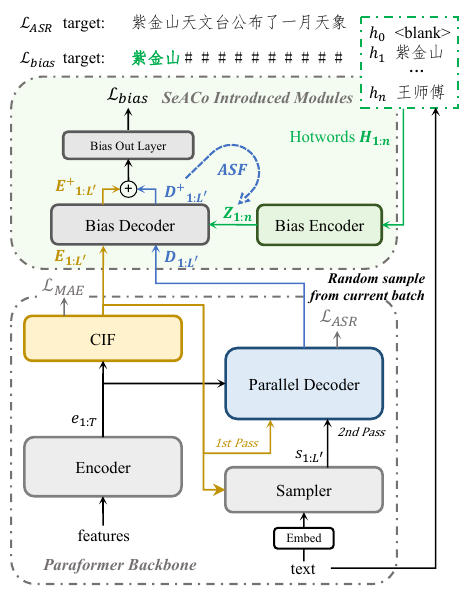}
    \caption{Illustration of SeACo-Paraformer.}\label{seaco}
    \vspace{-3mm}
\end{figure}

Taking the advantages and shortcomings of the previous works discussed in Section 1 into account, we propose a novel hotword customization system called Semantic-augmented Contextual-Paraformer~(SeACo-Paraformer). Based on Paraformer which is a strong ASR backbone, 3 modules are introduced as illustrated in Fig~\ref{seaco}. Intuitively, SeACo-Paraformer utilize the modeling characteristic of CIF predictor to conduct hotword prediction, while the CIF output and parallel decoder output are sent to bias decoder separately.

Formally, considering an speech feature $\mathbf{x}_{1:T}$ and corresponding text $\mathbf{y}_{1:L}$, we retain the CIF output $\mathbf{E}_{1:L^{'}}$ and parallel decoder hidden state~(before output layer) $\mathbf{D}_{1:L^{'}}$ in Paraformer inference:
\begin{equation}
\begin{aligned}
\mathbf{e}_{1: T} &=\operatorname{Encoder}\left(\mathbf{x}_{1: T}\right) \\
\mathbf{E}_{1: L^{'}} &=\operatorname{CIF}(\mathbf{e}_{1: T}) \\
\mathbf{s}_{1:L^{'}} &=\operatorname{Sampler}\left(\mathbf{E}_{1: L^{'}}, \operatorname{EMB}\left(\mathbf{y}_{1:L}\right)\right) \\
\mathbf{D}_{1:L^{'}} &=\operatorname{Parallel Decoder}\left(\mathbf{s}_{1:L^{'}}; \mathbf{e}_{1: T}\right)
\end{aligned}
\end{equation}
Then $n$ hotwords are randomly sampled out of batches of $\mathbf{y}_{1:L}$ with batch size $bs$, denoted as $\mathbf{H}_{1:n}$. We use 4 hyper-parameter here to control the sampling process: $r_b$ for the ratio of batches to conduct sampling,  the forward of the other batches will be conducted with a default hotword \textit{$\langle blank\rangle$}; $r_u$ is similar with $r_b$ but in utterance level inside an active batch, the average number of hotwords sampled for active batch is $r_u \times bs + 1$~(one for the default hotword); $l_{min}$ and $l_{max}$ for the minimum and maximum lengths of sampled hotwords. 

The character sequences in hotword list are then embedded with bias encoder which contains a embedding layer~(parameter shared with ASR embedding) and an LSTM layer:
\begin{equation}
\begin{aligned}
\mathbf{Z}_{1:n} = \operatorname{LSTM}\left(\operatorname{EMB}\left(\mathbf{H}_{1:n}\right)\right)
\end{aligned}
\end{equation}
$\mathbf{Z}_{1:n}\in R^{n \times d}$ is unsqueezed and repeated on the 0th dimension for batch calculation. Then comes to the main part of SeACo-Paraformer. Inside bias decoder, the bias information of hotwords is introduced to acoustic embedding $\mathbf{E}_{1:L^{'}}$ and decoder hidden state $\mathbf{D}_{1:L^{'}}$ through the attention mechanism:
\begin{equation}
\begin{aligned}
\mathbf{D^+}_{1:L^{'}} &= \operatorname{MultiHeadedAttention}\left(\mathbf{D}_{1:L^{'}}, \mathbf{Z}_{1:n}, \mathbf{Z}_{1:n}\right)\\
\mathbf{E^+}_{1:L^{'}} &= \operatorname{MultiHeadedAttention}\left(\mathbf{E}_{1:L^{'}}, \mathbf{Z}_{1:n}, \mathbf{Z}_{1:n}\right)
\end{aligned}
\end{equation}
Bias decoder is composed of several multi-headed attention and feed-forward layer. With biased acoustic embedding and biased decoder hidden state, the biased probabilities $\mathbf{P}_{ASR_{1:L^{'}}}$ can be obtained with an output layer. Note that an additional token (counted as \#, means no-bias) is appended to the ASR output vocabulary to mark non-hotword position outputs.
\begin{equation}
\mathbf{P_{b_i}}=\operatorname{BiasOutLayer}\left(\mathbf{D^+}_{1:L^{'}}+\mathbf{E^+}_{1:L^{'}}\right)
\end{equation}
Given the biased probabilities $\mathbf{P}_{b_{1:L^{'}}}$, the bias-related parameters   can be updated with the hotword-position-aware criterion in which labels in non-hotword positions are replaced by \#~(as shown as $L_{bias}$ in Fig~\ref{seaco}).

With a well-trained Paraformer model freezing, we enable contextualization of hotwords for an ASR system by introducing bias out layer, bias decoder and bias encoder, and training them with randomly sampled hotwords and their corresponding targets. Notably, the training of bias-related parameters is separate from the ASR training, thereby allowing for the use of specialized hotword data~(e.g. low frequency language phrases) and training strategies without affecting the general ASR performance. 

%Compared to the vanilla CIF-ColDec\cite{han2021cif} model, SeACo-Paraformer is based on the Paraformer model, which is a strong NAR ASR backbone system, and its bias decoder takes more complex information as input. Moreover, SeACo-Paraformer uses a different inference strategy and enables filtering of the hotword list with attention score inside the bias decoder, as explained below.

\subsection{Inference and Auxiliary Technics}
For $i$-th step in SeACo-Paraformer inference with a given hotword list, we get the final merged probabilities for contextualized ASR as
\begin{equation}
\mathbf{P}_{\mathbf{m}_{\mathbf{i}}}=\left\{\begin{array}{l}
\mathbf{P}_{\mathbf{A S R}_{\mathbf{i}}}, \text { if } \arg \max \mathbf{P}_{\mathbf{b}_{\mathbf{i}}}=\textit{$\langle no-bias\rangle$} \\
\lambda \mathbf{P}_{\mathbf{b}_{\mathbf{i}}}+(1-\lambda) \mathbf{P}_{\mathbf{A S R}}, \text { otherwise }
\end{array}\right.
\end{equation}
When there is no hotword incoming or no hotword detected, SeACo-Paraformer uses $\mathbf{P}_{ASR_{i}}$ only. $\lambda$ is a tunable parameter to adjust the degree of trust bias decoder output. 

In practical applications, as the number of incoming hotwords expands, the performance of hotword activation decreases accordingly - it becomes difficult for the cross-attention inside bias decoder to establish correct connection between ASR decoder output $\mathbf{D}_{1:L^{'}}$ and large-sacle sparse hotword embedding $\mathbf{Z}_{1:n}$. For enabling SeACo-Paraformer to conduct hotword customization with large scale hotword list, we propose attention score filtering~(ASF) strategy. The bias decoder inference is conducted with full hotword list first to obtain the attention score matrix $\mathbf{A} \in R^{L \times n}$, where $L$ is the length of output tokens and $n$ is the number of hotwords. Then we sum up the scores of steps in $L$ and get the attention scores for each hotword. According to attention scores, we can pick the most active $k$ hotwords to conduct truly effective bias decoder inference. Comparing to the \textit{fine-grained contextual knowledge selection}~\cite{han2022improving}, our bias decoder is composed of multiple cross-attention layers and we found the score from the last layer most effective for filtering.

\section{Experiments}

\label{sec:typestyle}
\subsection{Data Introduction}

We conducted a series of experiments using large-scale internal industrial data - the random sampling based hotword modeling strategy requires sufficient diversity of data, otherwise it will easily overfit to limited semantic information. The total amount of training data we use is around 50,000 hours. For evaluating the performance of models in hotword customization as well as general ASR, we use test sets in different domains as shown in Table\ref{data}. For all of the hotword test sets, we distinguish part of hotwords as \textit{R1-hotwords}, which means that their recall rate counted in the recognition of the basic ASR model is lower than 40\%. These hotwords exhibit a higher level of recall difficulty, thereby effectively showing the customization capability of the models.
%\begin{enumerate}
\textit{Test-General} is a large collection of several general ASR test sets, covering various domains and scenarios.
\textit{Test-Commercial}, \textit{Test-Term} and \textit{Test-Entity} are three hotword test sets of different domain.% \textit{Test-Commercial} is about insurance, finance, automotive and etc. \textit{Test-Term} is composed of utterances about several subjects and corresponding term as hotwords. \textit{Test-Entity} is more general and its hotwords is almost name entities of places and person names.

To demonstrate \textbf{reproducible} experiment results and model comparison, we construct and \textbf{open source} two data sets for testing hotword customization based on the open-source Aishell-1 dataset~\cite{bu2017aishell}. First we use part-of-speech tagging~\cite{jiao2018LAC} to generate name entity list from transcription of Aishell-1 test and development sets. Then treating the name entities as hotwords, we use all of the R1-hotwords and a random part of the remaining ones to compose two hotword lists. Finally, \textit{Test-Aishell1-NE} and \textit{Dev-Aishell1-NE} are filtered out based on the hotword lists\footnote{Find the open source data sets, source codes, detailed experiment results and models at https://github.com/R1ckShi/SeACo-Paraformer}. 
%\end{enumerate}
\vspace{-3mm}
\begin{table}[htbp]
\centering
\caption{Test sets for customization and general ASR.}
\scalebox{0.94}{
\begin{tabular}{lccc}

\hline
                 & \#utt & \#hotword & \#R1-hotword \\ \hline
\textit{Test-Gerneral}    & 40603 & -         & -            \\
\textit{Test-Commercial}  & 2000  & 693       & 72           \\
\textit{Test-Term}        & 1639  & 969       & 258          \\
\textit{Test-Entity}      & 1308  & 231       & 54           \\
\textit{Test-Aishell1-NE} & 808   & 400       & 226          \\
\textit{Dev-Aishell1-NE}  & 1334  & 600       & 371          \\ \hline
\end{tabular}\label{data}}
\end{table}
\vspace{-6mm}

\subsection{Experimental Setup and Evaluation Metrics}

We implement the models with FunASR\footnote{https://github.com/alibaba-damo-academy/FunASR}.
The Paraformer model, Paraformer-CLAS model and the proposed SeACo-Paraformer model are \textbf{totally open source}\footnotemark[1] including source code, configuration files and models trained with industrial data.  Paraformer base model is composed of 50-layers SAN-M~\cite{gao2020san} encoder and 16-layer NAR Transformer decoder, with 2048 hidden units. Comparing to the models discussed in \cite{gao2023funasr}, the Paraformer-CLAS presented in this paper has undergone an upgrade and is guaranteed to possess a strong and finely-tuned customization baseline. As for SeACo-Paraformer, the model is trained with ASR backbone freezing using the same 50,000 hours training data for 6 epochs. We set sampling ratio $r_u$ and $r_b$ to 0.75, $l_{min}=2$ and $l_{max}=8$. The introduced modules have following architectures: bias encoder is a 2-layer LSTM with 512 hidden size, bias decoder is a 4-layer Transformer decoder with 1024 hidden units. Each training batch has 6,000 speech frames~($\times 16$ GPU).

\begin{table*}[h]
\centering
\caption{Experiment results over all test sets.}
\includegraphics[width=0.94\linewidth]{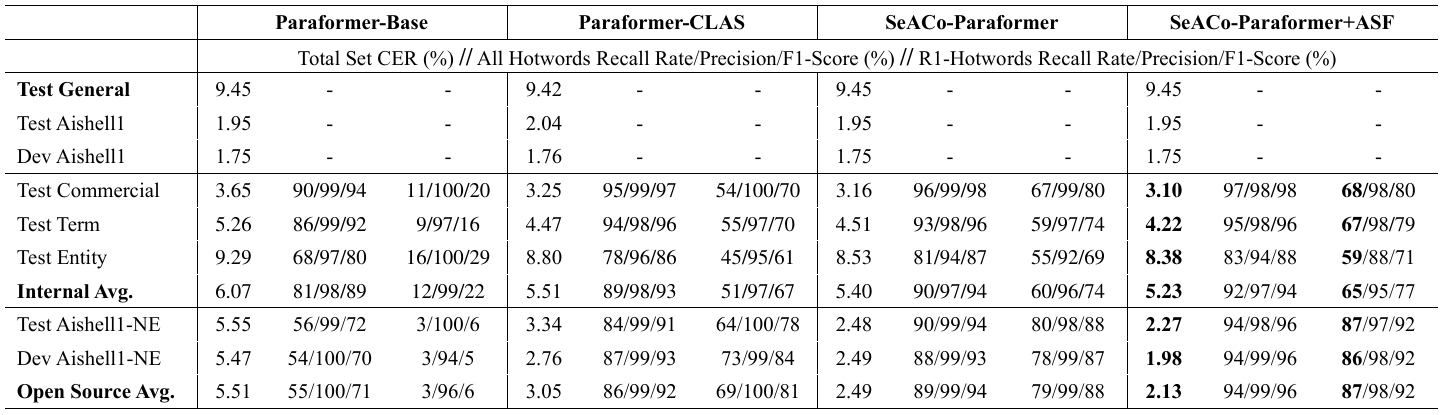}
\vspace{-3mm}
\label{res1}
\end{table*}

Models are compared according to 3 metrics: character error rate~(CER) over total test sets, recall rate/precision/F1 score~(R/P/F) of the all hotwords in average and R/P/F of R1-hotwords in average. For more strict evaluation, we only consider a hotword to be successfully recalled when predicted entirety, so we exclude the calculation of error rates in hotword positions.

\subsection{Results}
Table~\ref{res1} shows the experiment results over all test sets of the four models. Note that we use $\lambda=1.0$ for Equation~(5) in inference and $k=50$ for ASF. In the general test, Paraformer-CLAS demonstrated similar CER to the Paraformer-Base model. As the SeACo-Paraformer was trained with a frozen ASR backbone, the four models exhibited minimal differences in general ASR performance. In the internal hotword customization test, CLAS was shown to be an efficient method, with the average recall of R1-hotwords increasing from 12\% to 51\%. However, SeACo-Paraformer still outperformed CLAS with a recall rate of 60\%. The addition of ASF results in a further enhancement, the recall rate increased to 65\%. And SeACo-Paraformer achieved a 5\% relative CER reduction compared to Paraformer-CLAS. We get similar results in open-source test sets while Paraformer-CLAS, SeACo-Paraformer and SeACo-Paraformer+ASF get 69\%, 79\% and 87\% recall rates respectively. The proposed method achieves relative 30.2\% CER reduction and 26.1\% recall improvement comparing to Paraformer-CLAS.

\begin{figure}[htbp]
    \centering
    \includegraphics[width=0.92\linewidth]{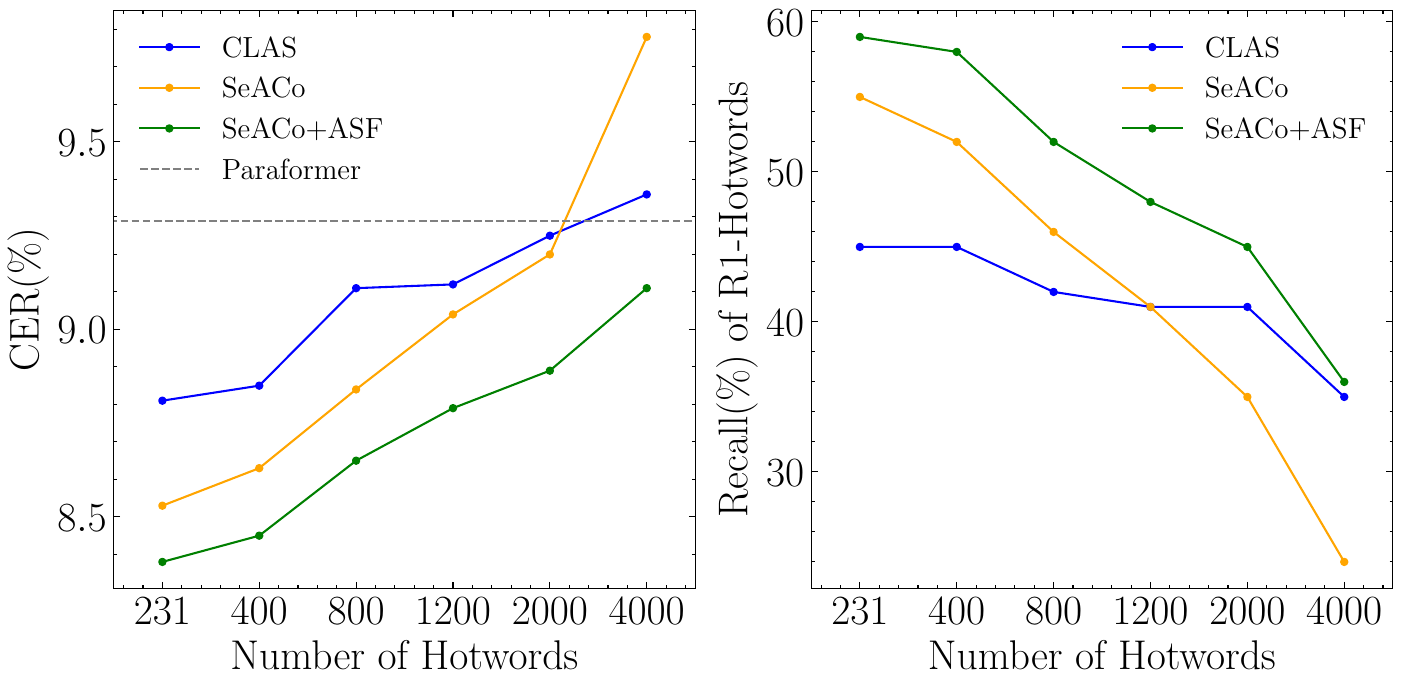}
    \vspace{-4mm}
    \caption{Results of increasing hotword number experiments.}
    \label{res2}
\end{figure}

The experiment results of increasing hotword number are shown in Fig.~\ref{res2}. \textit{Test-Entity} has 231 hotwords in total and we enlarge it with irrelevant hotwords to size of 400, 800, 1200, 2000 and 4,000. The curves in the figure indicate that the performance of all three models deteriorates as the hotword list expands. In terms of CER, only SeACo-Paraformer+ASF achieves a reduction in CER when the list expands to 4000 hotwords. From the perspective of recall rate, the SeACo-Paraformer model performs on par with with CLAS when the number of hotwords reaches 1200. However, by adopting ASF, the proposed model maintains its advantage over CLAS.

\begin{table}[]
\caption{Ablation study results (CER and R1-Hotword Recall)}
\centering
\scalebox{0.9}{
\begin{tabular}{lllll}
\hline
                 & \textit{Test -C} & \textit{Test -T} & \textit{Test-E} & Avg.    \\ \hline
Seaco-Paraformer & 3.16/67          & 4.51/59          & 8.53/55         & 5.40/60 \\
A1: merge first & 3.28/65          & 4.72/61          & 8.67/53         & 5.56/60 \\
A2: $\mathbf{E^+}$ only    & 3.86/64          & 5.29/62          & 8.66/58         & 5.94/61 \\
A3: $\mathbf{D^+}$ only   & 3.25/63          & 4.59/50          & 8.60/51         & 5.49/55 \\ \hline
\end{tabular}\label{res3}}
\vspace{-4mm}
\end{table}

Table~\ref{res3} shows the results of ablation study over calculation of bias decoder.
Focus on Equation~(3), there are several ways to introduce bias information before bias output layer, merging $\mathbf{E}_{1:L^{'}}$ and $\mathbf{D}_{1:L^{'}}$ before MHA rather thar after also makes sense~(A1). Besides, it's essential to conduct ablation experiment for using $\mathbf{E}_{1:L^{'}}$ or $\mathbf{D}_{1:L^{'}}$ only~(A2 and A3). The result indicates that using CIF output alone to calculate hotword loss leads to higher recall but worse in CER, which means that predicting hotword with acoustic embedding only is not a stable method and decoder output is indispensable. On the other hand, conducting MHA with $\mathbf{E}_{1:L^{'}}$ and $\mathbf{D}_{1:L^{'}}$ twice and separately is better than merging them first.

\section{Analysis}
\begin{figure}[htbp]
    \centering
    \includegraphics[width=0.95\linewidth]{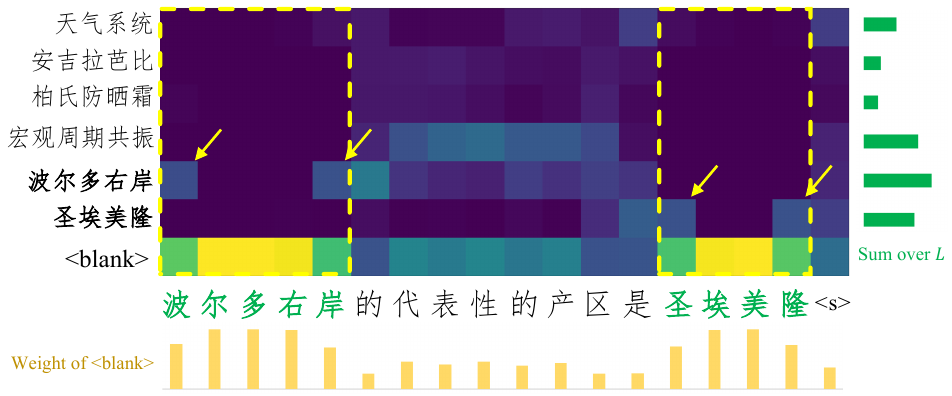}
    \caption{Bias decoder attention score matirx $\mathbf{A} \in R^{L \times n}$.}
    \label{attn}
    \vspace{-5mm}
\end{figure}

\vspace{-2mm}
The attention in bias decoder plays the most critical role in hotword customization, in this section we go deep into the attention score matrix. Figure~\ref{attn} reveals how the bias decoder establishes the connection between hotwords and semantic information~(attention score matrix of the last layer in bias decoder), but the phenomenon contradicts our initial hypothesis. In this case SeACo-Paraformer predicts two hotwords correctly. Two yellow dotted boxes depict the attention pattern upon successful matches, revealing that semantic information selectively attends to non-blank phrases solely when relevant hotwords initiate and conclude (indicated by yellow arrows). This phenomenon occurs in almost all utterances and is not an isolated case. Besides, we found the attention scores of \textit{$\langle blank\rangle$} at each step imply the probability of hotwords to appear, which is different from \cite{han2021cif}. The cumulative scores across all steps for each hotword reflect its likelihood of occurrence, which constitutes the fundamental concept of ASF.

\section{Conclusion}
In this paper we propose a flexible and effective hotword customization model based on NAR ASR backbone called SeACo-Paraformer. A series of experiments with industrial data proved that the proposed model outperforms CLAS which is a classic hotword customization solution, and the ASF strategy for inference is proved an effective method to improve ASR accuracy and hotwords recall rate. Besides, we make further analysis over the performance with increasing number of hotwords, ablation study of bias decoder calculation and the attention score matrix. In the future, we will focus on large-scale incoming hotword list and different bias encoder structure.

\bibliographystyle{IEEE}
\bibliography{strings,refs}

\begin{thebibliography}{10}

\bibitem{graves2013speech}
Alex Graves, Abdel-rahman Mohamed, and Geoffrey Hinton,
\newblock ``Speech recognition with deep recurrent neural networks,''
\newblock in {\em 2013 IEEE international conference on acoustics, speech and
  signal processing}. IEEE, 2013, pp. 6645--6649.

\bibitem{chan2016listen}
William Chan, Navdeep Jaitly, Quoc Le, and Oriol Vinyals,
\newblock ``Listen, attend and spell: A neural network for large vocabulary
  conversational speech recognition,''
\newblock in {\em 2016 IEEE international conference on acoustics, speech and
  signal processing (ICASSP)}. IEEE, 2016, pp. 4960--4964.

\bibitem{vaswani2017attention}
Ashish Vaswani, Noam Shazeer, Niki Parmar, Jakob Uszkoreit, Llion Jones,
  Aidan~N Gomez, {\L}ukasz Kaiser, and Illia Polosukhin,
\newblock ``Attention is all you need,''
\newblock {\em Advances in neural information processing systems}, vol. 30,
  2017.

\bibitem{rao2017exploring}
Kanishka Rao, Ha{\c{s}}im Sak, and Rohit Prabhavalkar,
\newblock ``Exploring architectures, data and units for streaming end-to-end
  speech recognition with rnn-transducer,''
\newblock in {\em 2017 IEEE Automatic Speech Recognition and Understanding
  Workshop (ASRU)}. IEEE, 2017, pp. 193--199.

\bibitem{moritz2020streaming}
Niko Moritz, Takaaki Hori, and Jonathan Le,
\newblock ``Streaming automatic speech recognition with the transformer
  model,''
\newblock in {\em ICASSP 2020-2020 IEEE International Conference on Acoustics,
  Speech and Signal Processing (ICASSP)}. IEEE, 2020, pp. 6074--6078.

\bibitem{gao2020universal}
Zhifu Gao, Shiliang Zhang, Ming Lei, and Ian McLoughlin,
\newblock ``Universal asr: Unifying streaming and non-streaming asr using a
  single encoder-decoder model,''
\newblock {\em arXiv preprint arXiv:2010.14099}, 2020.

\bibitem{zhou2018multilingual}
Shiyu Zhou, Shuang Xu, and Bo~Xu,
\newblock ``Multilingual end-to-end speech recognition with a single
  transformer on low-resource languages,''
\newblock {\em arXiv preprint arXiv:1806.05059}, 2018.

\bibitem{karita2019comparative}
Shigeki Karita, Nanxin Chen, Tomoki Hayashi, Takaaki Hori, Hirofumi Inaguma,
  Ziyan Jiang, Masao Someki, Nelson Enrique~Yalta Soplin, Ryuichi Yamamoto,
  Xiaofei Wang, et~al.,
\newblock ``A comparative study on transformer vs rnn in speech applications,''
\newblock in {\em 2019 IEEE Automatic Speech Recognition and Understanding
  Workshop (ASRU)}. IEEE, 2019, pp. 449--456.

\bibitem{yu2021boundary}
Fan Yu, Haoneng Luo, Pengcheng Guo, Yuhao Liang, Zhuoyuan Yao, Lei Xie,
  Yingying Gao, Leijing Hou, and Shilei Zhang,
\newblock ``Boundary and context aware training for cif-based
  non-autoregressive end-to-end asr,''
\newblock in {\em 2021 IEEE Automatic Speech Recognition and Understanding
  Workshop (ASRU)}. IEEE, 2021, pp. 328--334.

\bibitem{gao2022paraformer}
Zhifu Gao, Shiliang Zhang, Ian McLoughlin, and Zhijie Yan,
\newblock ``Paraformer: Fast and accurate parallel transformer for
  non-autoregressive end-to-end speech recognition,''
\newblock {\em INTERSPEECH}, 2022.

\bibitem{aleksic2015improved}
Petar Aleksic, Cyril Allauzen, David Elson, Aleksandar Kracun, Diego~Melendo
  Casado, and Pedro~J Moreno,
\newblock ``Improved recognition of contact names in voice commands,''
\newblock in {\em 2015 IEEE International Conference on Acoustics, Speech and
  Signal Processing (ICASSP)}. IEEE, 2015, pp. 5172--5175.

\bibitem{hall2015composition}
Keith Hall, Eunjoon Cho, Cyril Allauzen, Francoise Beaufays, Noah Coccaro,
  Kaisuke Nakajima, Michael Riley, Brian Roark, David Rybach, and Linda Zhang,
\newblock ``Composition-based on-the-fly rescoring for salient n-gram
  biasing,''
\newblock 2015.

\bibitem{pundak2018deep}
Golan Pundak, Tara~N Sainath, Rohit Prabhavalkar, Anjuli Kannan, and Ding Zhao,
\newblock ``Deep context: end-to-end contextual speech recognition,''
\newblock in {\em 2018 IEEE spoken language technology workshop (SLT)}. IEEE,
  2018, pp. 418--425.

\bibitem{han2021cif}
Minglun Han, Linhao Dong, Shiyu Zhou, and Bo~Xu,
\newblock ``Cif-based collaborative decoding for end-to-end contextual speech
  recognition,''
\newblock in {\em ICASSP 2021-2021 IEEE International Conference on Acoustics,
  Speech and Signal Processing (ICASSP)}. IEEE, 2021, pp. 6528--6532.

\bibitem{huang2023contextualized}
Kaixun Huang, Ao~Zhang, Zhanheng Yang, Pengcheng Guo, Bingshen Mu, Tianyi Xu,
  and Lei Xie,
\newblock ``Contextualized end-to-end speech recognition with contextual phrase
  prediction network,''
\newblock {\em arXiv preprint arXiv:2305.12493}, 2023.

\bibitem{hu2019phoneme}
Ke~Hu, Antoine Bruguier, Tara~N Sainath, Rohit Prabhavalkar, and Golan Pundak,
\newblock ``Phoneme-based contextualization for cross-lingual speech
  recognition in end-to-end models,''
\newblock {\em arXiv preprint arXiv:1906.09292}, 2019.

\bibitem{jain2020contextual}
Mahaveer Jain, Gil Keren, Jay Mahadeokar, Geoffrey Zweig, Florian Metze, and
  Yatharth Saraf,
\newblock ``Contextual rnn-t for open domain asr,''
\newblock {\em arXiv preprint arXiv:2006.03411}, 2020.

\bibitem{sathyendra2022contextual}
Kanthashree~Mysore Sathyendra, Thejaswi Muniyappa, Feng-Ju Chang, Jing Liu,
  Jinru Su, Grant~P Strimel, Athanasios Mouchtaris, and Siegfried Kunzmann,
\newblock ``Contextual adapters for personalized speech recognition in neural
  transducers,''
\newblock in {\em ICASSP 2022-2022 IEEE International Conference on Acoustics,
  Speech and Signal Processing (ICASSP)}. IEEE, 2022, pp. 8537--8541.

\bibitem{han2022improving}
Minglun Han, Linhao Dong, Zhenlin Liang, Meng Cai, Shiyu Zhou, Zejun Ma, and
  Bo~Xu,
\newblock ``Improving end-to-end contextual speech recognition with
  fine-grained contextual knowledge selection,''
\newblock in {\em ICASSP 2022-2022 IEEE International Conference on Acoustics,
  Speech and Signal Processing (ICASSP)}. IEEE, 2022, pp. 8532--8536.

\bibitem{graves2006connectionist}
Alex Graves, Santiago Fern{\'a}ndez, Faustino Gomez, and J{\"u}rgen
  Schmidhuber,
\newblock ``Connectionist temporal classification: labelling unsegmented
  sequence data with recurrent neural networks,''
\newblock in {\em Proceedings of the 23rd international conference on Machine
  learning}, 2006, pp. 369--376.

\bibitem{dong2020cif}
Linhao Dong and Bo~Xu,
\newblock ``Cif: Continuous integrate-and-fire for end-to-end speech
  recognition,''
\newblock in {\em ICASSP 2020-2020 IEEE International Conference on Acoustics,
  Speech and Signal Processing (ICASSP)}. IEEE, 2020, pp. 6079--6083.

\bibitem{gao2023funasr}
Zhifu Gao, Zerui Li, Jiaming Wang, Haoneng Luo, Xian Shi, Mengzhe Chen, Yabin
  Li, Lingyun Zuo, Zhihao Du, Zhangyu Xiao, et~al.,
\newblock ``Funasr: A fundamental end-to-end speech recognition toolkit,''
\newblock {\em arXiv preprint arXiv:2305.11013}, 2023.

\bibitem{bu2017aishell}
Hui Bu, Jiayu Du, Xingyu Na, Bengu Wu, and Hao Zheng,
\newblock ``Aishell-1: An open-source mandarin speech corpus and a speech
  recognition baseline,''
\newblock in {\em 2017 20th conference of the oriental chapter of the
  international coordinating committee on speech databases and speech I/O
  systems and assessment (O-COCOSDA)}. IEEE, 2017, pp. 1--5.

\bibitem{jiao2018LAC}
Zhenyu Jiao, Shuqi Sun, and Ke~Sun,
\newblock ``Chinese lexical analysis with deep bi-gru-crf network,''
\newblock {\em arXiv preprint arXiv:1807.01882}, 2018.

\bibitem{gao2020san}
Zhifu Gao, Shiliang Zhang, Ming Lei, and Ian McLoughlin,
\newblock ``San-m: Memory equipped self-attention for end-to-end speech
  recognition,''
\newblock {\em arXiv preprint arXiv:2006.01713}, 2020.

\end{thebibliography}

\end{document}